\documentclass[journal=jacsat,manuscript=article]{achemso}

\usepackage[version=3]{mhchem} % Formula subscripts using \ce{}

\author{Arvind Saini}
\author{Rajiblochan Sahoo}
\author{Rajarshi Chakrabarti}
\email{rajarshi@chem.iitb.ac.in}
\affiliation[ChemistryIITB]
{Department of Chemistry, Indian Institute of Technology Bombay, Mumbai-400076,
India.}
\author{Sayantan Dutta}
\email{sayantan.dutta@iitb.ac.in}
\affiliation[ChemicalIITB]
{Department of Chemical Engineering, Indian Institute of Technology Bombay, Mumbai-400076,
India.}
\title[An \textsf{achemso} demo]
{A theoretical framework for investigating the role of stiffness heterogeneity in structure and dynamics of flexible polymer}
\begin{document}

%\begin{tocentry}
%\begin{center}
 %   \includegraphics[height=4.5 cm]{toc_enrtry.pdf}\end{center}
%\end{tocentry}

\begin{abstract}
  \noindent 
  Heteropolymers are ubiquitous in both synthetic systems, such as block copolymers, and biological macromolecules, including proteins and nucleic acids. Beyond their chemical composition, these polymers often exhibit spatial variations in physical properties. For instance, in biopolymers such as chromatin, stiffness heterogeneity arises from inherent molecular features as well as extensile or contractile active forces. In this letter, we develop a theoretical framework that extends the physics of flexible polymers, a widely used tool to describe biopolymer dynamics, to incorporate spatially varying stiffness. Using this approach, we specifically analyze the structure and dynamics of flexible heteropolymers with periodic stepwise stiffness profiles. We find that stiffness heterogeneity leads to qualitative deviations in dynamical observables such as mean squared displacement while also increasing structural anisotropy. Altogether, this framework provides a platform to interpret stiffness heterogeneity from experimental data especially for biopolymers 
  as well as to design heteropolymers with tailored structural and dynamic properties.
\end{abstract}
\section{Main Text}
%\section{Introduction}
 \noindent 
Heteropolymers are prevalent in both natural and synthetic systems. For example, block copolymers, which consist of distinct polymer segments arranged in a specific sequence, self-assemble into diverse morphologies and have applications ranging from generating thermally responsive polymer, soft lithography to drug delivery~\cite{feng2017block}. Similarly, many biological polymers, such as nucleic acids and proteins, are inherently heteropolymeric. Proteins are composed of different amino acids, while nucleic acids consist of distinct nucleotides. Beyond their chemical composition, the physical properties of polymer segments also vary along the chain\cite{zheng2023influence}, influencing their organization as well as biological function\cite{ricci2015chromatin,kepper2008nucleosome, girard2024heterogeneous,cao2024motorized,wakim2024physical,sandholtz2020physical,baldi2020beads}. For instance, heterochromatin, which is gene-poor, is more tightly packed than euchromatin, which is transcriptionally active, although they are part of the same polymer chain\cite{ashwin2019organization,mir2019chromatin,klemm2019chromatin}.

Among these physical properties, stiffness can also vary along the polymer chain. For example, a recent study revealed how stretchability of the segments vary along a chromatin chain along with other physical properties utilizing a data-driven coarse grained model\cite{kadam2023predicting}. This variablity of stiffness may arise from inherent molecular property as well as local stretching or compression of the polymer due to interaction with proteins such as transcription factors, RNA polymerases or topoisomerases at specific genomic locations\cite{racki2008atp, clapier2017mechanisms,mahajan2022euchromatin,eshghi2022symmetry, shin2023transcription,goychuk2023polymer}. Specifically, a recent work shows that the dynamics of a polymer with dipolar extensile or contractile active force along the chain can be effectively represented by a chain with reduced or enhanced stiffness\cite{chaki2023polymer}.

The physics of flexible polymers, particularly the Rouse model\cite{rouse1953theory, doi1988theory, khatri2007rouse}, has been instrumental in providing a theoretical foundation for describing the dynamics of biopolymers such as chromatin, whose total length far exceeds its persistence length. On one hand, these models consistently represent the chromatin dynamics in a broad range of length and timescale\cite{bajpai2023mesoscale,dutta2023leveraging}. On the other hand, Rouse models has been widely extended to consider relevant biophysical phenomena such as inherent viscoelasticity of the background\cite{weber2010bacterial,kailasham2021rouse}, active fluctuations\cite{samanta2016chain,chaki2019enhanced,goswami2022reconfiguration, ghosh2022active, chaki2023polymer, dutta2024effect,ubertini2024universal,winkler2020physics}, and formation of loops\cite{gabriele2022,sheng2004loop,polovnikov2023crumpled,yuan2024effect}. However, most existing models assume a constant stiffness across the polymer chain, represented by a uniform Kuhn length.
In this letter, we extend the theoretical approaches of flexible polymer dynamics by incorporating variable stiffness (Kuhn length) of the chain in the model to capture the effect of spatial heterogeneity of stiffness in the structure and dynamics of the polymer. We demonstrate the utility of this framework by analyzing how the structure and dynamics of the polymer chain with a periodic stepwise stiffness profile qualitatively and quantitatively deviates from a Rouse chain with uniform stiffness.

\begin{figure}[b!]
\centering
  \includegraphics{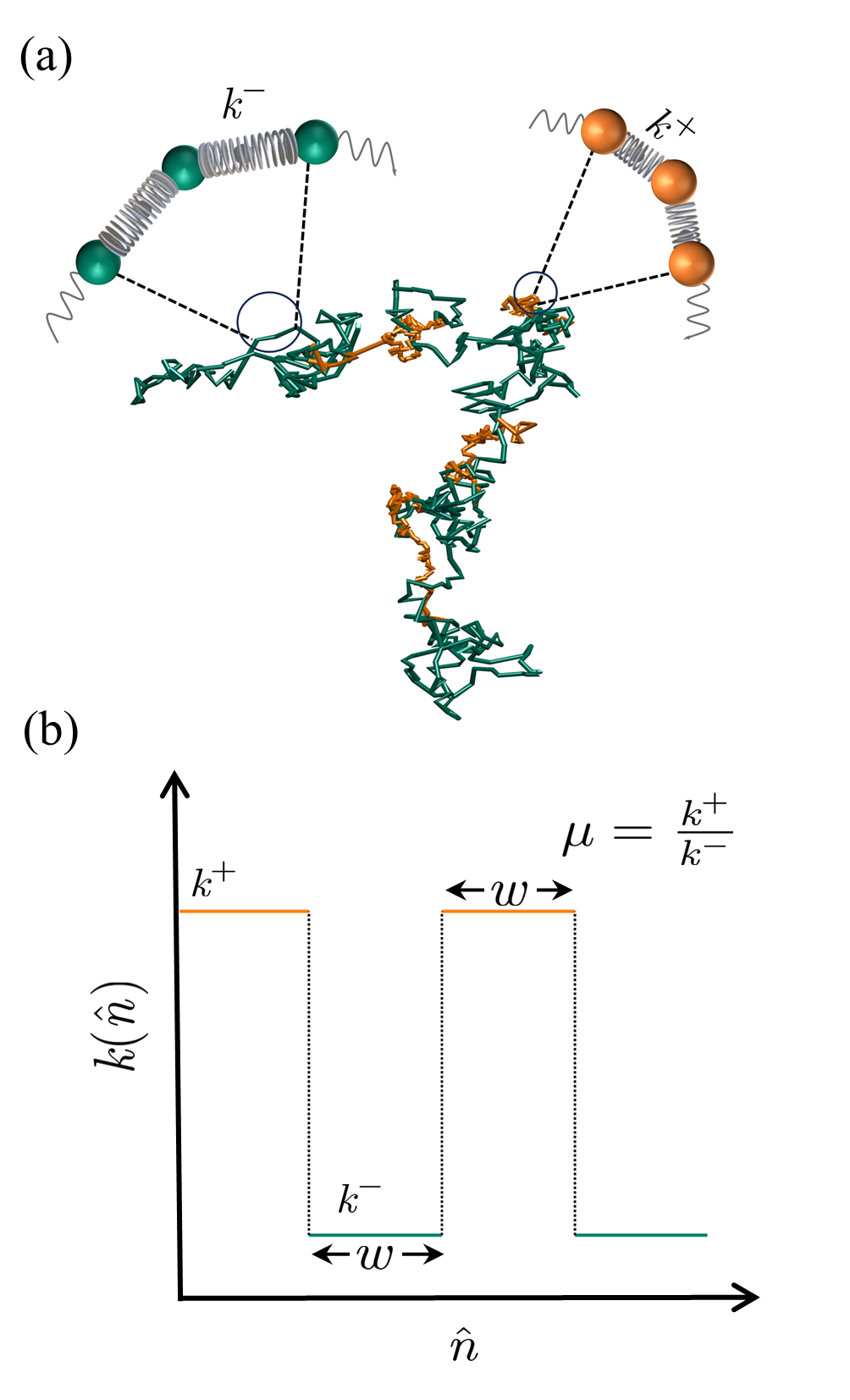}
  \caption{ (a) Conformation of a flexible polymer chain subject to heterogeneous stiffness, showing distinct regions of high stiffness ($k^+$, orange) and low stiffness ($k^-$, green). (b) 
  Relative stiffness of the segments as a function of segment position relative to the chain length (i.e $\hat n=n/N$). The domains with higher ($k^+$) and lower ($k^-$) stiffness alternate periodically. The parameters that characterize this functional form are $\mu = k^+/k^-$ and $w$, the half-width  of the periodic pattern.}
  \label{fig:Schematic}
\end{figure}

We describe the dynamics of a flexible polymer chain in a viscous medium, with variable stiffness along the chain (Fig.~\ref{fig:Schematic}(a)) by adapting the Rouse model. The polymer chain is described as a space curve $\Vec{r}(n,t)$, where $n$ represents the sequence of Kuhn segments, which runs from $0$ to $N$, the total number of Kuhn segments in the polymer. In this model, each segment evolves as a function of time $t$ according to the Langevin equation of motion:

\begin{equation}
\zeta \frac{\partial \Vec{r}(n,t)}{\partial t} = \frac{3 k_B T}{\langle b^2\rangle} \left( \frac{\partial }{\partial n}\left( k(n) \frac{\partial \Vec{r}(n,t)}{\partial n}\right) \right) + \Vec{f}^{\rm{B}}(n,t),
\label{EoM}
\end{equation}

\noindent where, $ \zeta $ is the viscous drag coefficient, and $ \Vec{f}^{\rm{B}}(n,t) $ is the instantaneous Brownian force acting on each  segment due to thermal fluctuations characterized by temperature $T$ and obeys the fluctuation dissipation theorem:
\begin{equation}
\langle \Vec{f}^{\rm{B}}(n,t) \Vec{f}^{\rm{B}}(n',t') \rangle = 2 k_B T \zeta \delta(n - n') \delta(t - t') \mathbf{I},
\label{Fluc_diss}
\end{equation}
where $ \mathbf{I} $ is the identity matrix\cite{doi1988theory}. We represent the heterogeneity in stiffness by defining a local Kuhn length $b(n)= \sqrt {\frac{\langle{b^2\rangle}}{k(n)}}$, where $\langle b^2\rangle$, is the mean-squared Kuhn length of the polymer chain and $k(n)$ is the relative stiffness of the chain, which necessitates
\begin{equation}
    \frac{1}{N}\int_0^N \frac{1}{k(n)} {\rm d}n =1,
\label{k_integral}
\end{equation} and defines the mean squared end-to-end distance $N\langle b^2 \rangle$, and mean Rouse time $\langle{\tau_R}\rangle = {\zeta N^2 \langle{b^2}\rangle}/{3 \pi^2 k_B T}$, which we  use as the characteristic length and timescale for the rest of the manuscript. The boundary conditions at both ends of the polymer chain reflect the force-free condition:
\begin{equation}
\frac{\partial \Vec{r}(n,t)}{\partial n} \Big|_{n = 0} = \frac{\partial \Vec{r}(n,t)}{\partial n} \Big|_{n = N} = 0
\label{bound_cond}
\end{equation}

In the physics of flexible polymers, the polymer configuration $ \Vec{r}(n,t) $ is often represented as a linear superposition of orthonormal modes $ \phi_p(n) $:

\begin{equation}
\Vec{r}(n,t) = \sum_{p=0}^{\infty} \Vec{X}_p(t) \phi_p(n),
\label{mode_decomp}
\end{equation}
 where, $ \Vec{X}_p(t) $ are the amplitudes of eigenmodes $ \phi_p(n) $ of different eigenfunctions associated with Eq.~\ref{EoM}, which are defined as: 

\begin{equation}
\phi_p(n)=\left\{\begin{array}{cc}1 & p=0 \\ \sqrt{2} \cos \left(\frac{n p \pi}{N}\right) & p>0 .\end{array}\right.   
\end{equation}

 Projecting the modified Langevin equation onto these normal modes, we derive the equation of motion of amplitude of the $p^{th}$ eigenmode as (Supplementary Material note A):

\begin{equation}
\frac{d\Vec{X}_p(t)}{dt} + \frac{1}{\langle{\tau_{R}}\rangle} \sum_{q=0}^{\infty} A_{pq} \Vec{X}_q(t) = \frac {\Vec{f}_p^{\rm{B}}(t)}{N\zeta},
\label{mode_EOM}
\end{equation}

where, each element $A_{pq}$ is determined by the stiffness profile of the chain as:
\begin{align}
A_{pq} = 2pq\int_0^1 k(\hat{n}) \sin\left( p \pi \hat{n}\right) \sin\left( q \pi \hat{n}\right) d\hat{n},
\label{A_def}
\end{align}
with, $\hat n=n/N$.

 Next, using Eq.~\ref{Fluc_diss} and using Eq.~\ref{mode_EOM}, we derive the time correlation between the mode amplitudes,  $C_{pq}(t)=\langle {X}_p(t) {X}_q(0) \rangle$, as:
\begin{equation}
    \mathbf{C}(\tau) = \frac{N b^{2}}{\pi^{2}} \mathbf{A}^{-1} e^{\mathbf{-A}\tau},
    \label{C_def}
\end{equation}
where, time is non-dimesionalized as $\tau = \frac{t}{<\tau_{R}>}$ and the elements of the matrix $\mathbf{C}(\tau)$ and $\mathbf{A}$ are $C_{pq}(\tau)$, and $A_{pq}$ respectively. The mode-mode correlation is specifically instrumental in connecting the stiffness heterogeneity to different observables that quantify the structure and dynamics of the polymer, which we will explore further in this letter.

We systemically explore the effect of heterogeneity of stiffness utilizing the developed framework, focusing  on a specific form of $k(n)$ (Fig.~\ref{fig:Schematic} (b)). In this functional form, stiffness varies in a periodic stepwise fashion as a function of the sequence of segments, i.e., we study a heteropolymer composed of segments with high stiffness ($k^{+}$) and a low   stiffness($k^{-}$)  . For most of the manuscript, we consider polymers with equal fractions of high and low stiffness segments. This leads to $\frac{1}{k^{+}}+\frac{1}{k^{-}}=2$ as a consequence of Eq.~\ref{k_integral}. In the following part of the manuscript, we will specifically explore the role of two parameters: (i) $\mu=k^{+}/k^{-}$, the stiffness ratio, which uniquely determines the values of $k^{+}$ as $(\mu+1)/2$ and $k^{-}$ as $(\mu+1)/2\mu$ and (ii) $w$, the width of the segment with continuous stiffness relative to the total length of the chain.

\begin{figure}[t]
\centering
  \includegraphics{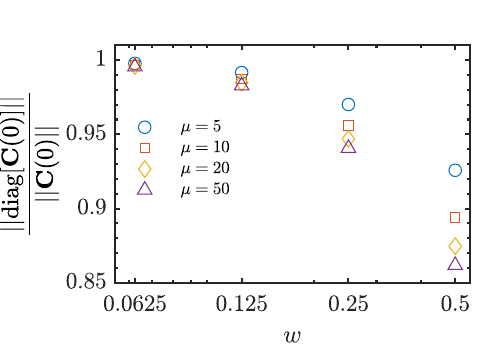}
  \caption{The ratio of the norm of the diagonal elements of correlation matrix $\mathbf C$ to the norm of the entire matrix
  as a function of width of the uniform stiffness region ($w$) and the stiffness ratio $\mu$.}
  \label{fig:Ratio_of_C_pq}
\end{figure}

We first examine the effect of heterogeneity of stiffness in the mode-mode correlation matrix $\mathbf C$ itself. For a Rouse polymer with uniform stiffness, cross-mode correlations are absent, leading to a purely diagonal correlation matrix. However, as evident from Eq.~\ref{A_def} and ~\ref{C_def}, the cross-mode correlations appear in a polymer with heterogeneous stiffness. To quantify this effect, we measure the diagonality of $\mathbf C (0)$ by computing the ratio of the norm of its diagonal elements to the norm of the entire matrix. Our analysis reveals that as the stiffness ratio $\mu$ and the width of individual regions $w$ increase, this ratio deviates further from 1, indicating a growing contribution from cross-mode correlations. In the following part of the letter, we will investigate how these contributions alter the dynamics and structure of the polymer compared to a Rouse polymer with uniform stiffness.

\begin{figure*}[t!]
\centering
   \includegraphics[width=0.95\linewidth]{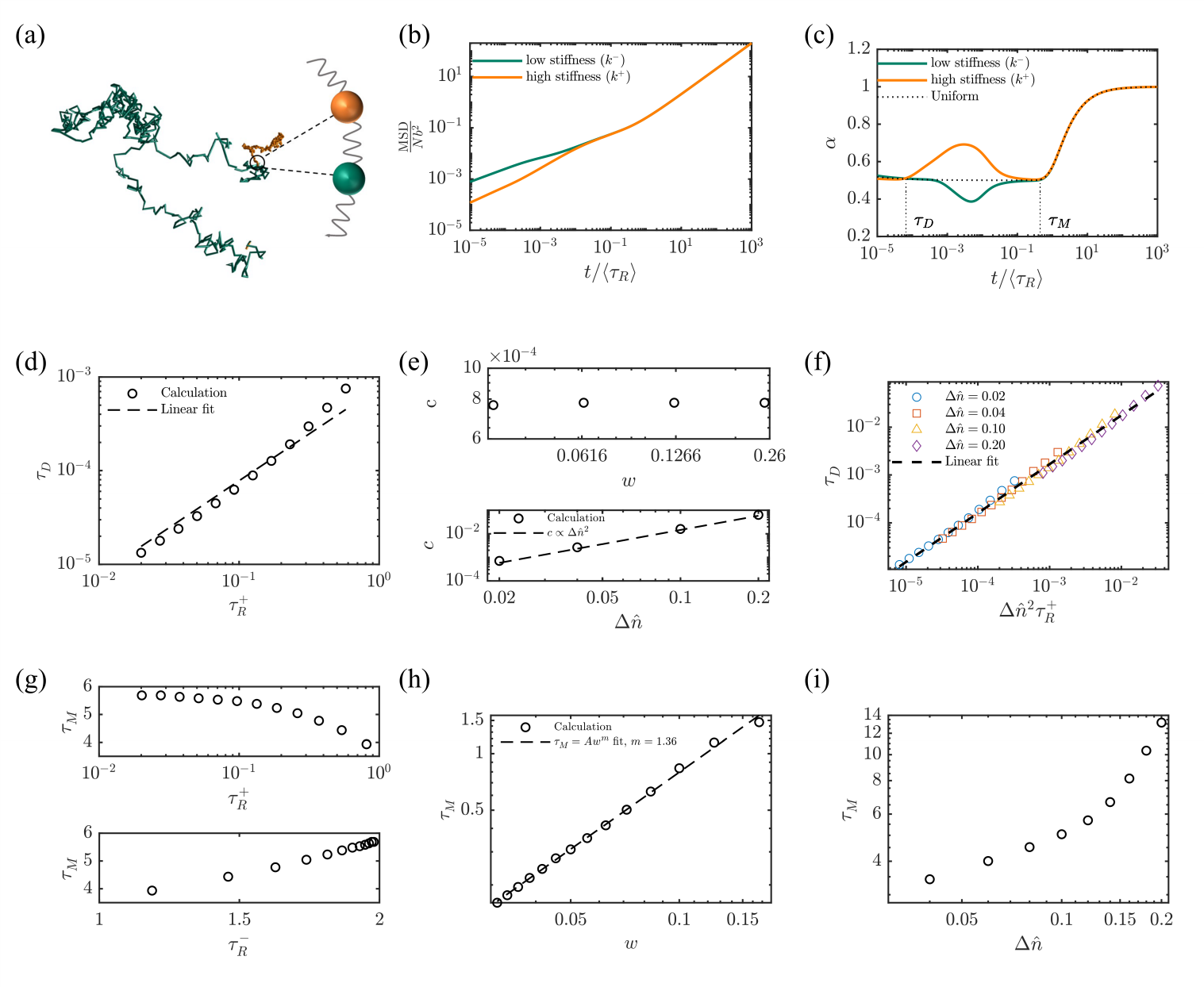}
\caption{(a) Schematic of a polymer with tagged segments from high (orange) and low (green) stiffness regions.  (b) Mean squared displacement (MSD) of tagged monomers from high (orange) and low (green) stiffness regions. (c) Scaling exponent ($\alpha$) of MSD as a function of time for segments from high (orange solid line) and low (green solid line) stiffness  regions, along with a segment from a uniform-stiffness Rouse chain (dotted line). Vertical dotted lines indicate deviation time ($\tau_D$) and merging time ($\tau_M$). (d) Relationship between $\tau_D$ and $\tau_R^+$, with the best-fit line ($\tau_D = c\tau_R^+$) shown as a dotted line. 
    (e) Best-fit intercept $c$ as a function of domain width ($w$) and segment separation ($\Delta \hat{n}$).  (f)$\tau_D$ as a function of $\Delta \hat{n}^2 \tau_R^+$ for different segment separations, with a best-fit linear trend (dotted line).  
    (g) Dependence of $\tau_M$ on $\tau_R^+$ and $\tau_R^-$.  
    (h) Relationship between $\tau_M$ and domain width ($w$). The dotted line represents  best fit power law fit. ($\tau_M = A w^m$), where $m = 1.36$. (i) Dependence of $\tau_M$ on segment separation ($\Delta \hat{n}$).  
    }
    \label{fig:dynamics}
\end{figure*}

 We quantify the influence of stiffness heterogeneity on polymer dynamics by computing the mean squared displacement (MSD) of individual Kuhn segments. We would like to note that, this data is experimentally accessible for biopolymers (such as chromatin) from imaging of fluorescent proteins tagged to a specific segment.  In Supplementary Material Note B, we show that the expression of the MSD of the $n$th segment reads as,

\begin{equation}
\text{MSD}(n, t) =  \frac{2 N \langle b^2 \rangle \tau}{\pi^2} + 2 \sum_{p=1}^{\infty} \sum_{q=1}^{\infty} \phi_p(n) \phi_q(n) \left[ C_{pq}(0) - C_{pq}(t) \right] 
\end{equation}

where, $ \tau = t / \langle{\tau_R}\rangle $. The indices $p$ and $q$ go over all orthonormal modes, and $\phi_p$ and $\phi_q$ are the eigenfunctions representing them. In this part, we specifically compare the MSD of two segments located symmetrically around the middle of the chain ($ n = N/2 $), separated by $ \Delta n $ segments (Fig.~\ref{fig:dynamics} (a)). We define $ k(n) $ such that the chain's center always coincides with the boundary between high- and low-stiffness regions. As a result, one of these segments belongs to the high-stiffness region, while the other lies in the low-stiffness region. 

As illustrated in Fig.~\ref{fig:dynamics}(b), at short timescales, the segment in the high stiffness region is more constrained and exhibits slower dynamics than the segment in the low-stiffness region. However, at longer timescales, their mean squared displacements (MSD) converge, following the same trajectory.  To further investigate their dynamics, we compare the local MSD exponent (Fig.~\ref{fig:dynamics}(c)),  
$\alpha = \frac{d \ln(\mathrm{MSD}(n,t))}{d \ln(t)}$ as a function of time for both segments, as well as for a segment at the same position from a uniform-stiffness Rouse chain with  Rouse time equal to the mean Rouse time of the heterogeneous chain.  At very short times, both segments exhibit subdiffusive behavior with  $\alpha \approx 0.5$, resembling the behavior of segments of two independent Rouse chains of different stiffness. At intermediate times, the exponents deviate: the high-stiffness segment shows $\alpha > 0.5$, while the low-stiffness segment has $\alpha < 0.5$. At long times, once the MSDs merge, the exponents also become identical. This behavior arises because the high-stiffness segment at short time has a slower dynamics in comparison the low-stiffness segment. To follow the same (i.e, the center-of-mass trajectory) at long times, the high-stiffness segment speeds up while the low-stiffness segment slows down in the intermediate timescale, leading to the observed deviations. 

The observed behavior remains consistent across different values of $ \mu $ and $ w $ (Supplementary Fig. 1), though the timescales of transitions between regimes depend on them. We focus on two key timescales: the deviation time ($\tau_D$), when the exponents begin to diverge, and the merger time ($\tau_M$), when both the exponents and MSDs converge (Fig.~\ref{fig:dynamics}(c)).  We investigate whether these timescales correlate with the characteristic relaxation times of the high ($\tau_R^+ = 1 / k^+ $) and low-stiffness ($\tau_R^- = 1 / k^- $) regions, the segment width ($w$), or their separation ($\Delta \hat  n=\Delta n/N$). Our analysis reveals that $ \tau_D $ scales linearly with $ \tau_R^+ $ (Fig.~\ref{fig:dynamics}(d)), remains independent of the segment width (Fig.~\ref{fig:dynamics}(e), top), and is proportional to the square of the segment separation (Fig.~\ref{fig:dynamics}(e), bottom). These findings suggest that $ \tau_D $ proportional to $ \Delta \hat{n}^2 \tau_R^+ $, the stress communication time between the domain boundary to the specific segment, which we confirm in Fig.~\ref{fig:dynamics}(f). This implies that segments behave as part of a uniform-stiffness (Rouse) chain until stress from a region of different stiffness reaches them. This time scale is shorter for the high-stiffness segment and it deviates earlier from Rouse behavior, denoting $ \tau_D $ (Fig.~\ref{fig:dynamics}(c)). 

A similar analysis of $ \tau_M $ shows a weak dependence on $ \tau_R^+ $ but a strong correlation with $ \tau_R^- $ (Fig.~\ref{fig:dynamics}(g)). $ \tau_M $ depends significantly on both the domain width ($w$) and separation $ (\Delta \hat{n}) $ (Fig.~\ref{fig:dynamics}(h, i)). Specifically, we find a power-law scaling $ \tau_M \propto w^{1.36} $ (Fig.~\ref{fig:dynamics}(h)). In summary, $ \tau_M $ is controlled by the stress relaxation timescale of the low-stiffness region, as well as higher width of  high- and low-stiffness domains lead to longer merging times before both the segments recover Rouse behavior. We note that all the timescales reported in Fig.~\ref{fig:dynamics}(d)-(i) are non-dimensionlized with respect to the mean rouse time $\langle \tau_R \rangle$.

\begin{figure*}[t]
\centering
   \includegraphics[width=17.5 cm]{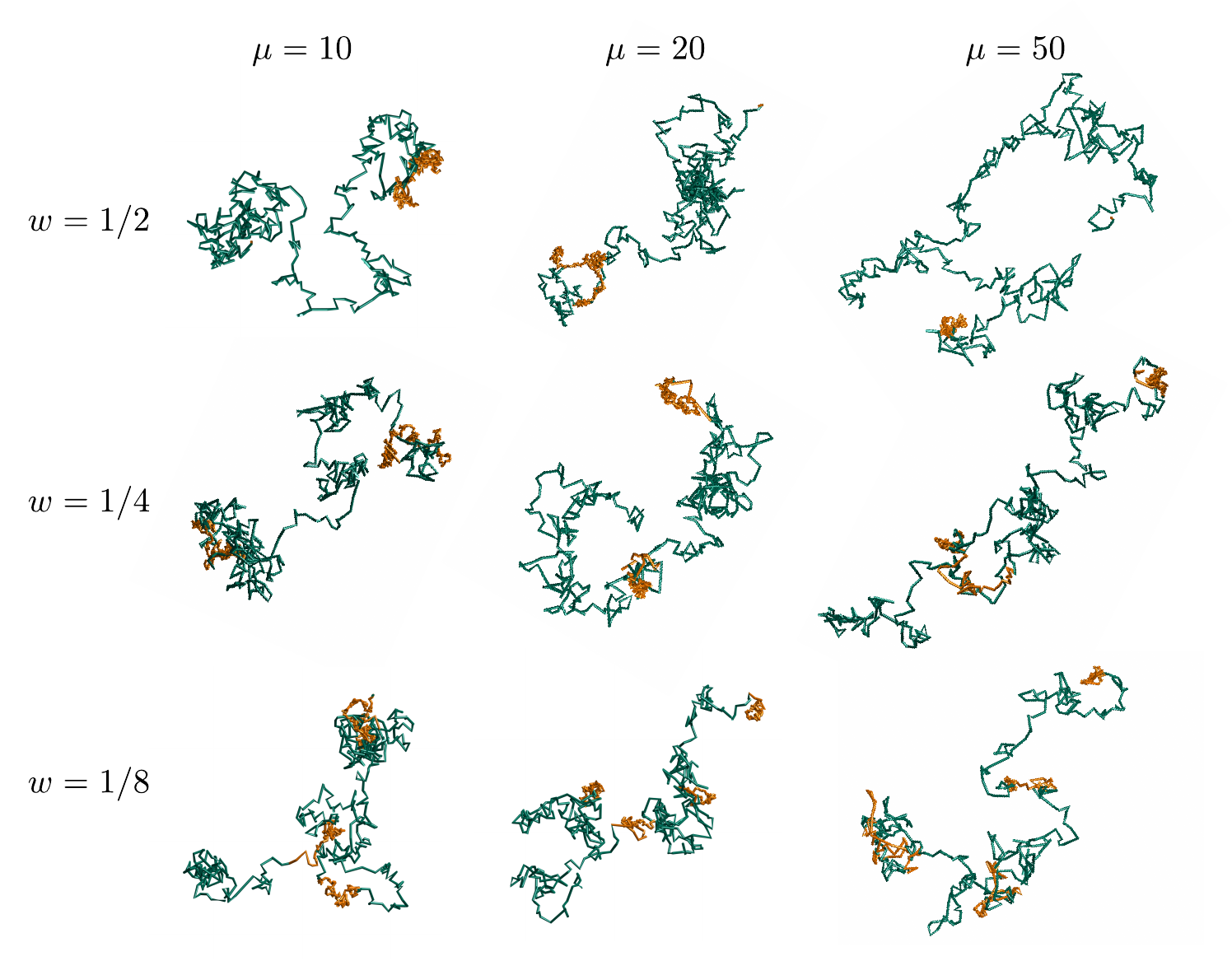}
\caption{Conformations of  polymers with same RMS Kuhn length but different ratio of stiffness ($\mu$) and width of domains of continuous stiffness ($w$) as illustrated in Fig.~\ref{fig:Schematic}(b). Orange and green colors represent segments with high and low stiffness, respectively. }
\label{fig:Conformations}
\end{figure*}

Next, we examine the conformation of polymers with heterogeneous stiffness. We begin with visualizing their equilibrium configurations. Using Eq.~\ref{C_def}, we compute the covariance matrix of the eigenmodes, $
\mathbf{C}(0) = \langle \mathbf{X} \mathbf{X}^{\rm T} \rangle$,
and generate eigenmodes \(\mathbf{X} = \{\vec{X}_p\}\) from a Gaussian distribution with zero mean and the same covariance matrix. We then apply Eq.~\ref{mode_decomp} to reconstruct equilibrium configurations \(\vec{r}(n)\) for different values of \(\mu\) and \(w\) (Fig.~\ref{fig:Conformations}).  As expected, regions with lower stiffness appear more swollen compared to stiffer regions. However, the overall size of the polymer remains nearly unchanged across different values of \(\mu\) and \(w\). Specifically, we compute the mean squared radius of gyration,  
$\langle R_g^2 \rangle = \sum_{p=1}^{\infty} C_{pp}(0)$
and find that it remains
$
\langle R_g^2 \rangle = N \langle b^2 \rangle /6
$ for all parameter values (Supplementary Fig. 2), indicating that global polymer dimensions are unaffected by stiffness heterogeneity.

%\subsection{Effect of stiffness heterogeneity on structural properties}

\begin{figure*}[t!]
\centering
   \includegraphics[width=\textwidth]{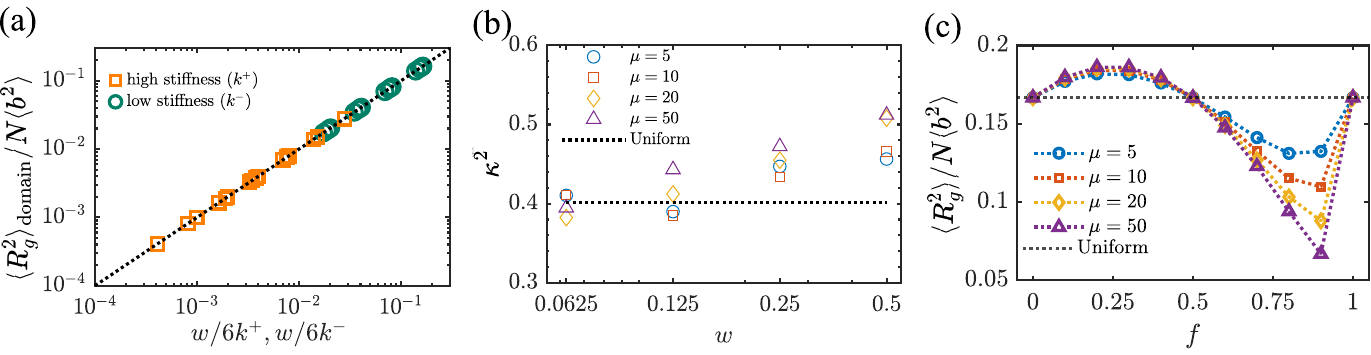}
\caption{(a) Domain wise mean squared radius of gyration $\langle R_g^2 \rangle_{\rm domain}$ plotted against the mean squared radius of gyration of the same domains, if they were part of an independent chain (i.e.$ w/6k^+ $ and $ w/6k^- $). The dotted line is a diagonal. (b) Anisotropy parameter ($ \kappa $) as a function of $ w $ for different values of $ \mu $. (c) Mean squared radius of gyration of the heteropolymers with same root mean squared Kuhn length $b$ but different fractions ($f$) of high stiffness ($k^{+}$) domain for diiferent values of stiffness contrast $\mu=k^{+}/k^{-}$. }
\label{fig:Structural_properties}
\end{figure*}

To investigate the local structure of the stiffness domains, we calculate the mean squared radius of gyration of individual domains, $\langle R_g^2 \rangle_{\rm domain}$, given by (Supplementary Material Note C):

\begin{equation}
\langle R_g^2 \rangle_{\rm domain} = \frac{1}{Nw} \sum_{p=1}^{\infty}\sum_{q=1}^{\infty} C_{pq}(0) \int_{N_1}^{N_2} \Big[ \phi_p \phi_q - \phi_p \overline{\phi_q} - \overline{\phi_p} \phi_q + \overline{\phi_p}  \overline{\phi_q} \Big]  dn,
\end{equation}
where, $N_1$ and $N_2$ denote the segment boundaries of a stiffness domain, and $\overline{\phi_p}$ and $\overline{\phi_q}$ are the mean values of the respective eigenfunctions over the domain. We find that, irrespective of $\mu$ and $w$, the mean squared radius of gyration of high and low stiffness domains follows $ \langle R_g^2 \rangle_{\rm high} = Nw \langle b^2 \rangle / 6k^+ $ and $ \langle R_g^2 \rangle_{\rm low} = Nw \langle b^2 \rangle / 6k^- $, respectively (Fig.~\ref{fig:Structural_properties}(a)). This indicates that each domain behaves as an independent polymer chain with its corresponding stiffness.

Since stiffness heterogeneity does not significantly impact global or local polymer size, we next compute the anisotropy factor $\kappa$, defined as:

\begin{equation}    
\kappa^2 = \frac{3}{2} \frac{\lambda_1^2+\lambda_2^2+\lambda_3^2}{\left( \lambda_1+\lambda_2+\lambda_3 \right)^2} - \frac{1}{2},
\end{equation}
where, $\lambda_1, \lambda_2,$ and $\lambda_3$ are the eigenvalues of the gyration tensor. The anisotropy factor quantifies deviations from a spherical shape, with $\kappa = 0$ indicating a perfect sphere and $\kappa = 1$ representing a rod-like structure. We note that a Rouse chain is inherently aspherical with $\langle \kappa^2 \rangle \sim 0.4$ in spite of absence of any inherent anisotropy.  We calculate $\langle \kappa^2 \rangle$ from 100 configurations similar to the ones shown in Fig.~\ref{fig:Conformations} for different parameter values.  For small $\mu$ and $w$, anisotropy values are close to those of a uniform-stiffness (Rouse) chain (Fig.~\ref{fig:Structural_properties}(b)). However, as $\mu$ and $w$ increase, anisotropy deviates significantly, shifting towards larger values, indicating increased elongation. This deviation is consistent with the relative fraction of  cross-mode correlations shown in Fig.~\ref{fig:Ratio_of_C_pq} for different $\mu$ and $w$.

Finally, we asked if the global structure is unaffected by stiffness heterogeneity for any form of $k(n)$. To answer that, we relax the constrain of equal proportion of low and high stiffness regions from the functional form of $k(n)$ shown in Fig.~\ref{fig:Schematic} (b) and analyze chains where a fraction $f$ of segments have high stiffness $k^{+}$, while the remaining $1-f$ have low stiffness $k^{-}$ (Supplementary Fig. 3). Interestingly, $\langle R_g^2 \rangle = N \langle b^2 \rangle / 6$ holds true only for uniform stiffness chains and chains with equal fractions of high and low stiffness segments. When $0 <f < 0.5$, $\langle R_g^2 \rangle > N \langle b^2 \rangle / 6$, and when $0.5 <f<1 $, $\langle R_g^2 \rangle < N \langle b^2 \rangle / 6$~(Fig.~\ref{fig:Structural_properties} (c)).  The effect becomes more prominent as $\mu=k^{+}/k^{-}$ increases. This suggests that in chains with an unequal fraction of stiffness domains, the minority domain is influenced by the majority domain. For example, in a chain dominated by high-stiffness segments, low-stiffness segments are more restricted than those of an independent low-stiffness chain.

In this Letter, we present an analytical framework that extends the physics of flexible polymers to investigate how stiffness heterogeneity influences polymer structure and dynamics. We specifically utilize this framework to systematically investigate the properties of polymers, where the stiffness of the chain varies in a periodical and stepwise manner along the chain~(Fig.~\ref{fig:Schematic}). We observe that unlike a Rouse chain with uniform stiffness, cross-mode correlations appear in these chains~(Fig.~\ref{fig:Ratio_of_C_pq}). Consequently, while segmental mean squared displacements (MSD) resemble those of a Rouse chain at very short and very long timescales, they are qualitatively different in the intermediate timescales~(Fig.~\ref{fig:dynamics}).  On the other hand, although the size of the chains (characterized by radius of gyration) are largely unaffected by the heterogenity of stiffness, the polymers adopt more anisotropic conformations~(Fig.~\ref{fig:Conformations},~\ref{fig:Structural_properties}).These deviations from Rouse-like behavior become more pronounced as the stiffness contrast (i.e. the ratio of stiffness) and the domain width of high and low stiffness regions increase.

We would like to note that this framework could be utilized for any stiffness profile besides the one presented here, as long as the integral in Eq.~\ref{A_def} converges. 
By leveraging this approach, synthetic polymers with tailored stiffness distributions can be designed to achieve specific structural and dynamic properties. On the other hand, inverse problem could be framed utilizing this framework to predict about the experimentally inaccessible local stiffness profile of bio-polymers such as chromatin from the experimentally accessible data such as mean squared displacement (MSD) of multiple tagged segments.  Since local stiffness often serves as a proxy for active forces that contract or extend the polymer, this framework also offers a way to probe underlying biological mechanisms. Furthermore, as it is based on the physics of flexible polymer, the framework can easily be extended to take into account hydrodynamic interaction as well as local active fluctuations in the future. Altogether, this framework enhances  the toolbox of physics of flexible polymer that connects inherent heterogeneity in the polymer chain to it's structure and dynamics.

\begin{acknowledgement}

A.S. thanks IIT Bombay for the fellowship. R.S. thanks IIT Bombay for the fellowship. R.C. acknowledges SERB,
India under the MATRICS scheme for funding (Project No.
MTR/2020/000230). S.D. thanks the IIT Bombay seed grant (Project no: RD/0523-IRCCSH0-022). The authors acknowledge the SpaceTime-2 supercomputing facility at IIT Bombay for the computing time.

\end{acknowledgement}

\begin{suppinfo}
Detailed derivations and supplementary figures are available in Supplementary Material. 
\end{suppinfo}

\bibliography{achemso-demo}

\clearpage
\section{Supplementary Material}
\setcounter{equation}{0} 
\subsection{A. Temporal correlation for normal modes of a flexible polymer subject to stiffness heterogeneity}
\noindent In this section, we will derive the correlation of the normal modes \(p\) and \(q\) in time \(t\) for a flexible polymer with hetrogeneity in stiffness. We start our analysis by writing down the overdamped Langevin equation for the same: 
\begin{align}
\zeta \frac{\partial \Vec{r}(n,t)}{\partial t} = \frac{3 k_B T}{\langle b^2\rangle} \left( \frac{\partial }{\partial n}\left( k(n) \frac{\partial \Vec{r}(n,t)}{\partial n}\right) \right) + \Vec{f}^{\rm{B}}(n,t) 
\end{align}
subject to the boundary condition:
\begin{equation}
\frac{\partial \Vec{r}(n,t)}{\partial n} \Big|_{n = 0} = \frac{\partial \Vec{r}(n,t)}{\partial n} \Big|_{n = N} = 0
\label{bound_cond}
\end{equation}.

The eigenmodes satisfying the Boundary conditions are:

\begin{align}
\phi_p(n)=\left\{\begin{array}{cc}1 & p=0 \\ \sqrt{2} \cos \left(\frac{n p \pi}{N}\right) & p>0 .\end{array}\right.   
\end{align}

\noindent Taking the inner product of both sides of the Eq. 1 with the eigenmode \( \phi_p(n) \). 

\begin{equation}
\begin{aligned}
 \int_0^N\phi_p(n) \zeta \frac{\partial \Vec{r}(n,t)}{\partial t} \, dn &= \frac{3k_{B}T}{\langle b^2\rangle} \int_0^N \phi_p(n) \left( k(n) \frac{\partial^2 \Vec{r}(n,t)}{\partial n^2} \right) \,dn \\
&\quad + \frac{3k_{B}T}{\langle b^2\rangle} \int_0^N \phi_p(n)  \left( \frac{\partial k}{\partial n}\right) \left( \frac{\partial \Vec{r}(n,t)}{\partial n} \right) \, dn \\
&\quad + \int_0^N \phi_p(n) \Vec{f}^{\rm{B}}(n,t) \, dn
\end{aligned}
\end{equation}

\noindent Let's handle each term of RHS of Eq. 4 separately.

\noindent First Term of RHS of Eq. 4:
\begin{align}
\int_0^N \phi_p(n) \frac{3k_{B}T}{\langle b^2\rangle} k(n) \frac{\partial^2 \Vec{r}(n, t)}{\partial n^2} \, dn &= \frac{3k_{B}T}{\langle b^2\rangle} \int_0^N \phi_p(n) k(n) \frac{\partial^2 \Vec{r}(n, t)}{\partial n^2} \, dn \notag \\
&= \left. \frac{3k_{B}T}{\langle b^2\rangle} \left(\phi_p(n) k(n) \frac{\partial \Vec{r}(n, t)}{\partial n} \right|_0^N - \int_0^N \frac{\partial}{\partial n} \left( \phi_p(n) k(n) \right) \frac{\partial \Vec{r}(n, t)}{\partial n} \, dn \right) \notag \\
&= -\frac{3k_{B}T}{\langle b^2\rangle} \int_0^N \left( k(n) \frac{d\phi_p(n)}{dn} + \phi_p(n) \frac{dk(n)}{dn} \right) \frac{\partial \Vec{r}(n, t)}{\partial n} \, dn
\end{align}
\noindent First term in Eq. 5 vanishes due to boundary conditions.
Substituting eq. (5) in Eq, (4), we obtain:

\begin{align}
    \int_0^N\phi_p(n) \zeta \frac{\partial \Vec{r}(n,t)}{\partial t} \, dn = - \frac{3k_{B}T}{\langle b^2\rangle} \int_0^N k(n)  \left( \frac{\partial \phi_p(n) }{\partial n}\right) \left( \frac{\partial \Vec{r}(n,t)}{\partial n} \right) \, dn + \int_0^N \phi_p(n) \Vec{f}^{\rm{B}}(n,t) \, dn
\end{align}

\noindent The polymer configuration \( \Vec{r}(n, t) \) can be expressed as a sum of orthonormal modes:
\begin{align}
\Vec{r}(n, t) = \sum_{q=0}^{\infty} \Vec{X}_q(t) \phi_q(n)
\end{align}
where \( \Vec{X}_q(t) \) are the mode amplitudes and \( \phi_q(n) \) are the eigenfunctions.

Substituting \( \Vec{r}(n, t) \) into the LHS Integral of the Eq. 6, we obtain:
\begin{align}
\int_0^N \phi_p(n) \zeta \frac{\partial}{\partial t} \left( \sum_{q=0}^{\infty} \Vec{X}_q(t) \phi_q(n) \right) \, dn
\end{align}

\begin{align}
\int_0^N \phi_p(n) \zeta \sum_{q=0}^{\infty} \frac{d\Vec{X}_q(t)}{dt} \phi_q(n)
\, dn = \zeta \sum_{q=0}^{\infty} \frac{d\Vec{X}_q(t)}{dt} \int_0^N \phi_p(n) \phi_q(n) \, dn
\end{align}

\noindent The eigenfunctions  are orthogonal:
\begin{align}
\int_0^N \phi_p(n) \phi_q(n) \, dn = N\delta_{pq}
\end{align}
where \( \delta_{pq} \) is the Kronecker delta. Substitution of  Eq. (10) into Eq. (9) simplifies the left hand side of Eq. (6) as:
\begin{align}
\int_0^N \phi_p(n) \zeta \frac{\partial \Vec{r}(n, t)}{\partial t} \, dn = N\zeta \frac{d \Vec{X}_p(t)}{dt}
\end{align}
\noindent Next, we break the term $\frac{\partial \Vec{r}(n, t)}{\partial n}$ in orthonomal modes:

\begin{align}
\frac{\partial \Vec{r}(n, t)}{\partial n} = \sum_{q=0}^{\infty} \Vec{X}_q(t) \frac{d \phi_q(n)}{dn}
\end{align}
and substituting  this converts the RHS of Eq. (6) as,
\begin{align}
  -\frac{3k_{B}T}{\langle b^2\rangle} \left ( \sum_{q=0}^{\infty} \Vec{X}_q(t) \int_0^N k(n) \frac{d\phi_p(n)}{dn} \frac{d\phi_q(n)}{dn} \, dn \right) + \int_0^N \phi_p(n) \Vec{f}^{\rm{B}}(n,t) \, dn 
\end{align}

\noindent Combining Eq. 11 and Eq. 13, we get the equation of motion of normal mode amplitudes as, 
\begin{align}
N\zeta \frac{d\Vec{X}_p(t)}{dt} = -\frac{3k_{B}T}{\langle b^2\rangle} \left ( \sum_{q=0}^{\infty} \Vec{X}_q(t) \int_0^N k(n) \frac{d\phi_p(n)}{dn} \frac{d\phi_q(n)}{dn} \, dn \right) + \Vec{f_p^{\rm{B}}}(t),
\end{align}
where, 
$\Vec{f_p^{\rm{B}}}(t) = \int_0^N \vec{f^{\rm{B}}}(n, t) \phi_p(n) \, dn$

\noindent Now, we rephrase the Eq. (14) as,
\begin{align}
 \frac{d\Vec{X}_p(t)}{dt} &= - \frac{3 k_B T \pi^2}{\zeta \langle b^2 \rangle N^2} \sum_{q=0}^{\infty} A_{pq} \Vec{X}_q(t) + \frac{\Vec{f}_p^{\rm{B}}(t)}{N \zeta}
\end{align}
\noindent where,
\begin{align}
A_{pq} = 2pq\int_0^1 k(\hat{n}) \sin\left( p \pi \hat{n}\right) \sin\left( q \pi \hat{n}\right) d\hat{n},
\end{align}
with, $\hat n=n/N$.

\noindent We rewrite Eq. (15) using the mean Rouse time defined in main text $\langle\tau_{R}\rangle$ as, 
\begin{align}
    \frac{d\Vec{X}_p(t)}{dt} + \frac{1}{\langle \tau_{R}\rangle} \sum_{q=0}^{\infty} A_{pq} \Vec{X}_q(t) &= \frac {\Vec{f}_p^{\rm{B}}(t)}{N\zeta}
\end{align}
\noindent In matrix notation, this turns into 
\begin{align}
    \frac{d \mathbf{X}(\tau)}{dt} + \frac{1}{\langle \tau_{R}\rangle}  \mathbf{A} \mathbf{X}(\tau) &= \frac {\mathbf{f}(\tau)}{N\zeta}
\end{align}
\noindent where,
\[
\mathbf{X} = \begin{bmatrix}
\vec{X}_1 \\
\vec{X}_2 \\
\vdots \\
\vec{X}_\infty
\end{bmatrix}, \quad
\mathbf{A} = \begin{bmatrix}
A_{11} & A_{12} & \cdots & A_{1\infty} \\
A_{21} & A_{22} & \cdots & A_{2\infty} \\
\vdots & \vdots & \ddots & \vdots \\
A_{\infty1} & A_{\infty2} & \cdots & A_{\infty\infty}
\end{bmatrix}, \quad
\mathbf{f} = \begin{bmatrix}
\vec{f}_1 \\
\vec{f}_2 \\
\vdots \\
\vec{f}_\infty
\end{bmatrix}
\]

\noindent Now, the above equation is a linear first-order differential equation. So, the integrating factor for the latter is \( \text{I.F.} = e^{\mathbf{A}\tau} \), where 
\(\tau = \frac{t}{\tau_{R}}\)\\

\noindent Multiplying Eq. 18 by I.F., we get
\begin{align}
\frac{1}{\langle \tau_{R}\rangle}e^{\mathbf{A}\tau} \frac{d \mathbf{X}(\tau)}{d\tau} +\frac{1}{\langle \tau_{R}\rangle}  e^{\mathbf{A}\tau}   \mathbf{A} \mathbf{X}(\tau) &= e^{\mathbf{A}\tau} \frac {\mathbf{f}(\tau)}{N\zeta}
\end{align}

\noindent This simplifies to:
\begin{align}
\frac{1}{\langle \tau_{R}\rangle}\frac{d}{d\tau} \left( e^{\mathbf{A}\tau} \mathbf{X}(\tau) \right) &=e^{\mathbf{A}\tau} \frac {\mathbf{f}(\tau)}{N\zeta}
\end{align}

\noindent Integrating it both sides with respect to $\tau$:
\begin{align}
\frac{1}{\langle \tau_{R}\rangle}\int_{-\infty}^{\tau} \frac{d}{d\tau_1} \left( e^{\mathbf{A}\tau_1} \mathbf{X}(\tau_1) \right) d\tau_1 &= \int_{-\infty}^{\tau} e^{\mathbf{A}\tau_1} \frac {\mathbf{f}(\tau_1)}{N\zeta} d\tau_1
\end{align}

%\noindent The left-hand side simplifies to:
%\begin{align}
% e^{{\frac{A_{pq}}{\tau_{R}}t}} \Vec{X}_p(t) - \lim_{t \to -\infty}  e^{{\frac{A_{pq}}{\tau_{R}}t}} \Vec{X}_p(t)
%\end{align}

%\noindent Assuming that $\Vec{X}_p(t) \to 0$ as $t \to -\infty$, the limit term vanishes. So, 
\noindent The final solution for the above differential equation is:
\begin{align}
  \mathbf{X}(\tau) &= \frac{\langle\tau_{R}\rangle}{N \zeta} e^{\mathbf{-A}\tau} \int_{-\infty}^{\tau} e^{\mathbf{A}\tau_{1}} \mathbf{f}(\tau_{1}) d\tau_{1}
\end{align}

\noindent Similarly, we can find the solution at time 0 as,
\begin{align}
  \mathbf{X}(0) &= \frac{\langle\tau_{R}\rangle}{N \zeta} \int_{-\infty}^{0}  e^{\mathbf{A}\tau_{2}} \mathbf{f}(\tau_{2}) d\tau_{2}
\end{align} 

\noindent The correlation function \( \mathbf C(\tau) \) is:
\begin{align}
    \mathbf C(\tau) = \left\langle \mathbf{X}(\tau) \mathbf{X}^\mathrm{T}(0) \right\rangle
\end{align}

\noindent Substituting the solutions for \( \mathbf{X}(t) \) and \( \mathbf{X}(0) \):
\begin{align}
\mathbf C(\tau) &= \left\langle \left( \frac{\langle\tau_{R}\rangle}{N \zeta} e^{\mathbf{-A}\tau} \int_{-\infty}^{\tau} e^{\mathbf{A}\tau} \mathbf{f}(\tau_{1}) d\tau_{1} \right) \left( \frac{\tau_{R}}{N \zeta} \int_{-\infty}^{0}  e^{\mathbf{A}\tau} \mathbf{f}(\tau_{2}) d\tau_{2} \right)^\mathrm{T} \right\rangle
\end{align}

\noindent Using the correlation property of the random forces, we find the elements of the matrix
$\left\langle \mathbf{f}(\tau_1) \mathbf{f}^\mathrm{T}(\tau_2) \right\rangle = \mathbf{F}$
as,
\begin{align}
F_{pq} &= \left\langle f_p(t_1)\cdot  f_q(t_2)  \right\rangle \nonumber \\
&= \left\langle \int_0^N f^{\rm{B}}(n, t_1) \phi_p(n) \, dn \cdot  \int_0^N f^{\rm{B}}(n, t_2) \phi_q(n') \, dn' \right\rangle\nonumber\\
&= \int_0^N \int_0^N \left\langle f^{\rm{B}}(n, t_1) \cdot  f^{\rm{B}}(n, t_2) \right\rangle \phi_p(n) \phi_q(n') \, dn \, dn' \nonumber\\
&= 6 k_B T \zeta  \int_0^N \int_0^N \delta(t_1 - t_2) \delta(n - n') \phi_p(n) \phi_q(n') \, dn \, dn'\nonumber\\
&= 6 k_B T \zeta \delta(t_1 - t_2) \int_0^N \phi_p(n) \phi_q(n) \, dn \nonumber\\
&= 6 k_B T \zeta \delta(t_1 - t_2) N \delta_{pq} 
= \frac{6 k_B T  N \zeta }{\langle\tau_{R}\rangle}\delta(\tau_{1} - \tau_{2})\delta_{pq} 
\end{align}
\noindent So the matrix \(\mathbf{F} \) is, 
\begin{align}
   \left\langle \mathbf{f}(\tau_1) \mathbf{f}^\mathrm{T}(\tau_2) \right\rangle =  \frac{6 k_B T  N \zeta }{\langle\tau_{R}\rangle}\delta(\tau_{1} - \tau_{2})\mathbf{I}
\end{align}
\noindent Substituting  this into the expression for \( \mathbf C(\tau) \):
\begin{align}
\mathbf C(\tau) &= \frac{\langle\tau_{R}\rangle^{2}}{N^{2}} e^{\mathbf{-A}\tau} \int_{-\infty}^\tau \int_{-\infty}^0 e^{\mathbf{A}\tau_1} \frac{6  k_B T N \zeta \delta(\tau_1 - \tau_2) \mathbf{I}}{\tau_{R} \zeta^2}  e^{\mathbf{A}^\mathrm{T}\tau_2} \, d\tau_1 \, d\tau_2 \\
&= e^{\mathbf{-A}\tau} \frac{6 k_B T \langle\tau_{R}\rangle  \mathbf{I}}{N \zeta} \int_{-\infty}^\tau \int_{-\infty}^0 e^{\mathbf{A}\tau_1} \delta(\tau_1 - \tau_2) e^{\mathbf{A}^\mathrm{T}\tau_2} \, d\tau_1 \, d\tau_2
\end{align}

\noindent The delta function \( \delta(t_1 - t_2) \) simplifies the double integral to a single integral by enforcing \( \tau_1 = \tau_2 \) and sets the range of \(\tau_1 \) is from \( -\infty \) to \( 0 \) resulting in:

\begin{equation}
   \mathbf C(\tau)=e^{\mathbf{-A}\tau} \frac{6 k_B T \langle\tau_{R}\rangle  \mathbf{I}}{N \zeta}\int_{-\infty}^0 e^{(\mathbf{A} + \mathbf{A}^\mathrm{T}) \tau_1} \, d\tau_1 
\end{equation}

\noindent Also, $\mathbf A$ is a symmetric matrix (i.e. \( \mathbf{A} = \mathbf{A}^\mathrm{T} \)), and that further simplifies $\mathbf{C(\tau)}$ as:
\begin{align}
   \mathbf C(\tau)&= e^{\mathbf{-A}\tau} \frac{6 k_B T \langle\tau_{R}\rangle  \mathbf{I}}{N \zeta}\int_{-\infty}^0 e^{2\mathbf{A} \tau_1} \, d\tau_1\nonumber\\
   &= e^{\mathbf{-A}\tau} \frac{6 k_B T \langle\tau_{R}\rangle  \mathbf{I}}{N \zeta}(2 \mathbf A)^{-1} [\mathbf I-0]\nonumber \\
   &= \frac{6 k_B T\zeta N^2 \langle b^2 \rangle}{6 \pi^2 N \zeta K_B T}\mathbf A ^{-1}  e^{\mathbf{-A}\tau}\nonumber \\
   &= \frac{ N\langle b^2 \rangle}{\pi^2}\mathbf A ^{-1}  e^{\mathbf{-A}\tau}
\end{align}

\subsection{B. Mean Squared Displacement (MSD) of a tracer on a flexible polymer}

\noindent The Mean squared displacement (MSD) of a tracer located at segment \( n \) on a polymer is defined as:

\begin{equation}
\text{MSD}(n, t) = \langle [\vec{r}(n, t) - \vec{r}(n, 0)]^2 \rangle
\end{equation}

\noindent The position of a segment \( n \) on the polymer can be expressed as a sum of the center of mass (COM) position and the contributions from the normal modes:

\begin{equation}
\vec{r}(n, t) = \vec{r}_{\text{com}}(t) + \sum_{p=1}^{\infty} \vec{X}_p(t) \phi_p(n)
\end{equation}

\noindent Substituting the expression for \(\vec{r}(n, t)\) in Eq. 32, we obtain 

\noindent Expand this expression:

\begin{equation}
\begin{aligned}
\text{MSD}(n, t)  & =\left\langle \left[ \vec{r}_{\text{com}}(t) + \sum_{p=1}^{\infty} \vec{X}_p(t) \phi_p(n) - \vec{r}_{\text{com}}(0) - \sum_{p=1}^{\infty} \vec{X}_p(0) \phi_p(n) \right]^2 \right\rangle\\
&= \left\langle \left[ \vec{r}_{\text{com}}(t) - \vec{r}_{\text{com}}(0) \right]^2 \right\rangle + 2 \sum_{p=1}^{\infty} \phi_p(n) \left\langle \left[ \vec{r}_{\text{com}}(t) - \vec{r}_{\text{com}}(0) \right] \cdot \left[ \vec{X}_p(t) - \vec{X}_p(0) \right] \right\rangle \\
& \hspace {2.0 cm} + \left\langle \left[ \sum_{p=1}^{\infty} \vec{X}_p(t) \phi_p(n) - \sum_{p=1}^{\infty} \vec{X}_p(0) \phi_p(n) \right]^2 \right\rangle
\end{aligned}
\end{equation}

 The first term in Eq. 34 represents the MSD of the center of mass itself. The second term in Eq. 34 involves cross-correlations between the center of mass motion and the normal modes and the third term in Eq. 34 represents the contributions from the correlation between normal modes.

We begin by expanding the third term as:

\begin{align}
&\left\langle \left[ \sum_{p=1}^{\infty} \vec{X}_p(t) \phi_p(n) - \sum_{p=1}^{\infty} \vec{X}_p(0) \phi_p(n) \right]^2 \right\rangle\nonumber\\
&= \left\langle \sum_{p=1}^{\infty} \sum_{q=1}^{\infty} \left( \vec{X}_p(t)\phi_p(n) - \vec{X}_p(0)\phi_p(n) \right) \cdot \left( \vec{X}_q(t)\phi_q(n) - \vec{X}_q(0)\phi_q(n) \right) \right\rangle\nonumber\\
  &  = \sum_{p=1}^{\infty} \sum_{q=1}^{\infty} \phi_p(n) \phi_q(n) \left[ \left\langle \vec{X}_p(t) \cdot \vec{X}_q(t) \right\rangle - \left\langle \vec{X}_p(t) \cdot \vec{X}_q(0) \right\rangle - \left\langle \vec{X}_q(t) \cdot \vec{X}_p(0) \right\rangle + \left\langle \vec{X}_p(0) \cdot \vec{X}_q(0) \right\rangle \right]\nonumber\\
   & = \sum_{p=1}^{\infty} \sum_{q=1}^{\infty} \phi_p(n) \phi_q(n) \left[ 2C_{pq}(0) - C_{pq}(t) - C_{qp}(t) \right]
\end{align}
%\noindent Since \(\vec{X}_p(t)\) and \(\vec{X}_p_q(t)\) are normal mode amplitudes, their correlation is given by \(C_{pq}(t)\):

%\begin{equation}
%\left\langle \vec{X}_p(t) \vec{X}_q(0) \right\rangle = C_{pq}(t)
%\end{equation}
 Next, we analyze the center of mass contributions, we begin from the equation of motion stated in Eq. 1 and average it with respect to $n$
 \hspace{10 pt}
\begin{align}
\zeta\frac{\partial \Vec{r}(n,t)}{\partial t}& = \frac{3k_{B}T}{\langle b^2\rangle} \left( k(n) \frac{\partial^2 \Vec{r}(n,t)}{\partial n^2} + \left( \frac{\partial k}{\partial n}\right) \left( \frac{\partial \Vec{r}}{\partial n} \right)\right) + \Vec{f}^{\rm{B}}(n,t) \nonumber\\
\implies
\frac{\zeta}{N}\int_0^N\frac{\partial \Vec{r}(n,t)}{\partial t} \, dn &= \frac{3k_{B}T}{N b^2} \int_0^N\left( k(n) \frac{\partial^2 \Vec{r}(n,t)}{\partial n^2} + \left( \frac{\partial k}{\partial n}\right) \left( \frac{\partial \Vec{r}}{\partial n} \right)\right) \, dn + \frac{1}{N}\int_0^N\Vec{f}^{\rm{B}}(n,t) \, dn 
\end{align}

As the first integral of the right hand side vanishes, this equation simoplifies to
\begin{align}
    \frac{\partial \Vec{r}_\text{com}(t)}{\partial t} = \frac{1}{N \zeta}\int_0^N\Vec{f}^{\rm{B}}(n,t) dn
\end{align}
Integrating with time, we obtain:
\begin{align}
   \implies \Bigl[ \vec{r}_{\text{com}}(t) - \vec{r}_{\text{com}}(0) \Bigr] = \frac{1}{N\zeta}\int_0^t\int_0^N\Vec{f}^{\rm{B}}(n,t) \, dn \, dt
\end{align}
Next, we compute the MSD of the center of mass using the fluctuation-dissipation theorem as,
\begin{align}
    \Bigl \langle \Bigl[ \vec{r}_{\text{com}}(t) - \vec{r}_{\text{com}}(0) \Bigr]^2\Bigr \rangle &= \frac{1}{N^2\zeta^2}\int_0^t\int_0^N\int_0^t\int_0^N \Bigl \langle \Vec{f}^{\rm{B}}(n_1,t_1).\Vec{f}^{\rm{B}}(n_2,t_2) \Bigr \rangle \, dn_1 \, dt_1 \, dn_2 \, dt_2\nonumber\\
    &= \frac{3k_{B} T}{N^2\zeta^2}\int_0^t\int_0^N\int_0^t\int_0^N 2 \zeta \delta(n_1 - n_2) \delta(t_1 - t_2) \, dn_1 \, dt_1 \, dn_2 \, dt_2\nonumber\\
    & = \frac{3k_{B} T}{N^2} \bigl( 2Nt \bigr)
\end{align}

We rewrite the equation in terms of non-dimensional time using $\langle\tau_{R}\rangle = \frac{\zeta N^2 \langle b^2 \rangle}{3 \pi^2 k_B T}$ as,

\begin{align}
    \Bigl \langle \Bigl[ \vec{r}_{\text{com}}(\tau) - \vec{r}_{\text{com}}(0) \Bigr]^2\Bigr \rangle = \frac{3k_{B} T \langle\tau_{R}\rangle}{N^2} \bigl( 2N\tau \bigr)=\frac{2 N b^2\tau}{\pi^2} 
\end{align}

 Finally, we move on to calculate the correlations between the zeroth and  non-zero modes. Expressing $\mathbf{X}=\{\vec X_p\}$ and utilizing  the solution of the equation of motion of modal amplitudes stated in Eq. (22), we show that:
\begin{equation}
\begin{aligned}
&\hspace{10 pt}\left\langle \left[\vec{r}_{\text{com}}(\tau) - \vec{r}_{\text{com}}(0)\right] \cdot \left[\mathbf{X}(\tau) - \mathbf{X}(0)\right] \right\rangle\\
&= \left\langle \left[\vec{r}_{\text{com}}(\tau) - \vec{r}_{\text{com}}(0)\right] \cdot \frac{\langle\tau_{R}\rangle}{N\zeta} \left[e^{-\mathbf A \tau} \int_{-\infty}^\tau e^{\mathbf A \tau_1} \mathbf{f}(\tau_1) \, d\tau_1 \right.\right. \quad \left.\left. - \int_{-\infty}^0 e^{\mathbf A \tau_2} \mathbf{f}(\tau_2) \, d\tau_2\right] \right\rangle \\
&= \frac{\langle\tau_{R}\rangle}{N\zeta} \left\langle e^{-\mathbf A\tau} \left[\vec{r}_{\text{com}}(\tau) - \vec{r}_{\text{com}}(0)\right] \cdot \int_{-\infty}^\tau e^{\mathbf A \tau_1} \mathbf{f}(\tau_1) \, d\tau_1 \right\rangle \\
& \hspace{5 cm}- \frac{\langle\tau_{R}\rangle}{N\zeta} \left\langle \left[\vec{r}_{\text{com}}(\tau) - \vec{r}_{\text{com}}(0)\right] \cdot \int_{-\infty}^0 e^{\mathbf A \tau_2} \mathbf{f}(\tau_2) \, d\tau_2 \right\rangle\\
\end{aligned}
\end{equation}
\noindent Using Eq. 38 and expanding the the definition of $\mathbf f$, we can rewrite Eq. (41) as
\begin{equation}
    \begin{aligned}
&= \frac{\langle\tau_{R}\rangle}{N\zeta} \left\langle e^{-\mathbf A\tau}\int_0^\tau\int_0^N\Vec{f}^{\rm{B}}(n,\tau') \, dn \, d\tau'\cdot \int_0^N \int_{-\infty}^\tau e^{\mathbf A \tau_1} \vec{f}^{\rm{B}}(n_1,\tau_1) \mathbf \Phi(n_1) \, d\tau_1 \, dn_1 \right\rangle \\
&\hspace {5 cm} - \frac{\langle\tau_{R}\rangle}{N\zeta} \left\langle \int_0^\tau\int_0^N\Vec{f}^{\rm{B}}(n,\tau') \, dn \, d\tau' \cdot \int_0^N \int_{-\infty}^0 e^{\mathbf A \tau_2} \vec{f}^{\rm{B}}(n_2, \tau_2) \mathbf \Phi(n_2) \, d\tau_2 \, dn_2 \right\rangle,
    \end{aligned}
\end{equation}
where, $\mathbf \Phi (n)=\{\phi_p (n)\}$. Second term in Eq. 42 will be zero because Brownian force in the two integrals are time seperated and thus uncorrelated. Next, we expand the first term as, 

\begin{equation}
\begin{aligned}
    &\frac{\langle\tau_R\rangle}{N\zeta} \left\langle e^{-\mathbf A\tau}\int_0^\tau \int_0^N\Vec{f}^{\rm{B}}(n,\tau') \, dn \, d\tau' \cdot \int_0^N \int_{-\infty}^\tau e^{\mathbf A \tau_1} \vec{f}^{\rm{B}}(n_1,\tau_1) \mathbf \Phi (n_1) \, d\tau_1 \, dn_1 \right\rangle \\
     &= \frac{e^{-\mathbf A\tau} \langle \tau_R\rangle} {N\zeta}  \int_0^\tau\int_0^N \int_0^N \int_{-\infty}^\tau e^{\mathbf A \tau_1} \left\langle \Vec{f}^{\rm{B}}(n,\tau') \cdot     \vec{f}^{\rm{B}}(n_1,\tau_1)\right\rangle \mathbf \Phi (n_1) \, dn \, d\tau' \, d\tau_1 \, dn_1  \\
    &= \frac{e^{-\mathbf A\tau} \langle\tau_R\rangle 6k_BT \zeta}{N\zeta}  \int_0^\tau\int_0^N \int_0^N \int_{-\infty}^\tau e^{\mathbf A \tau_1} \delta(\tau' - \tau_{1}) \delta(n - n_1) \mathbf \Phi(n_1) \, dn \, d\tau' d\tau_1 \, dn_1  \\
    &= \frac{e^{-\mathbf A\tau} \langle \tau_R \rangle 6 K_B T }{N}  \int_0^\tau e^{\mathbf A \tau_1}  \, d\tau_1 \int_0^N    \mathbf \Phi (n_1)   dn_1  \\
\end{aligned}
\end{equation}

For the all eigenmodes $ \int_0^N    \phi_{p}(n)    \, dn = 0$, resulting in  $\mathbf \Phi (n_1)   dn_1=\mathbf 0$. Consequently for all $p$,
\begin{equation}
   \left\langle \left[\vec{r}_{\text{com}}(\tau) - \vec{r}_{\text{com}}(0)\right] \cdot \left[\vec{X}_p(\tau) - \vec{X}_p(0)\right] \right\rangle = 0 
\end{equation}
\noindent So, Final expression for mean squared displacement constitutes of only centre of mass contribution and contributions from cross-correlation of non-zero modes:
\begin{equation}
\begin{aligned}
\text{MSD}(n, t) = &  \frac{2 N b^2\tau}{\pi^2} + 2 \sum_{p=1}^{\infty} \sum_{q=1}^{\infty} \phi_p(n) \phi_q(n) \left[ C_{pq}(0) - C_{pq}(t) \right] 
\end{aligned}
\end{equation}

\subsection{C. Radius of Gyration}
The squared radius of gyration for an individual domain  is given by:
\begin{align}
\langle R_g^2 \rangle_{\rm domain} &= \left\langle (r_i - r_\text{com}^\text{domain})^2 \right\rangle,
\end{align}
where, $r_\text{com}^\text{domain}$ is the center of mass of the domain. In terms of normal mode the equation can be further expanded as,

\begin{equation}
\begin{aligned}
\langle R_g^2 \rangle_{\rm domain} &= \left\langle \frac{1}{Nw} \int_{N_1}^{N_2} \left[ \sum_{p=1}^{\infty} X_p \left(\phi_p(n) -\bar \phi_p(n)\right)  \right]^2 \, dn \right\rangle\\
&= \left\langle \frac{1}{Nw} \int_{N_1}^{N_2} \left[ \sum_{p=1}^{\infty} X_p \left(\phi_p(n) -\bar \phi_p(n)\right)  \right] \left[ \sum_{q=1}^{\infty} X_q\left(\phi_q(n) -\bar \phi_q(n)\right)  \right] \, dn \right\rangle\\
&= \frac{1}{Nw} \left\langle X_p X_q \right\rangle \int_{N_1}^{N_2} \left( \phi_p(n) - \overline{\phi_p} \right) \left( \phi_q(n) - \overline{\phi_q} \right) dn \\
&= \frac{1}{Nw} \sum_{p=1}^{\infty}\sum_{q=1}^{\infty} C_{pq}(0) \int_{N_1}^{N_2} \Big[ \phi_p \phi_q - \phi_p \overline{\phi_q} - \overline{\phi_p} \phi_q + \overline{\phi_p} \overline{\phi_q} \Big] \, dn,
\end{aligned}
\end{equation}
where,
\begin{itemize}
    \item $N_1$ and $N_2$ denotes the segment boundaries of a stiffness domain,
    \item $w$ is the width of the domain relative to chain length.
    \item $\phi_p(n)$ are eigenmodes as introduced in Eq. (3). 
    \item $\overline{\phi_p} = \frac{1}{Nw} \int_{N_1}^{N_2}\phi_p(n) \, dn$ is the average of $\phi_p(n)$ over the segment,
    \item $C_{pq}(0) = \langle X_p X_q \rangle$ represents the equilibrium covariance between eigenmode  amplitudes of mode $p$ and $q$.
\end{itemize}

\clearpage
\setcounter{figure}{0}  

\subsection*{Supplementary Figures}

\begin{figure*}[h!]
\centering
  \includegraphics[width=0.95\linewidth]{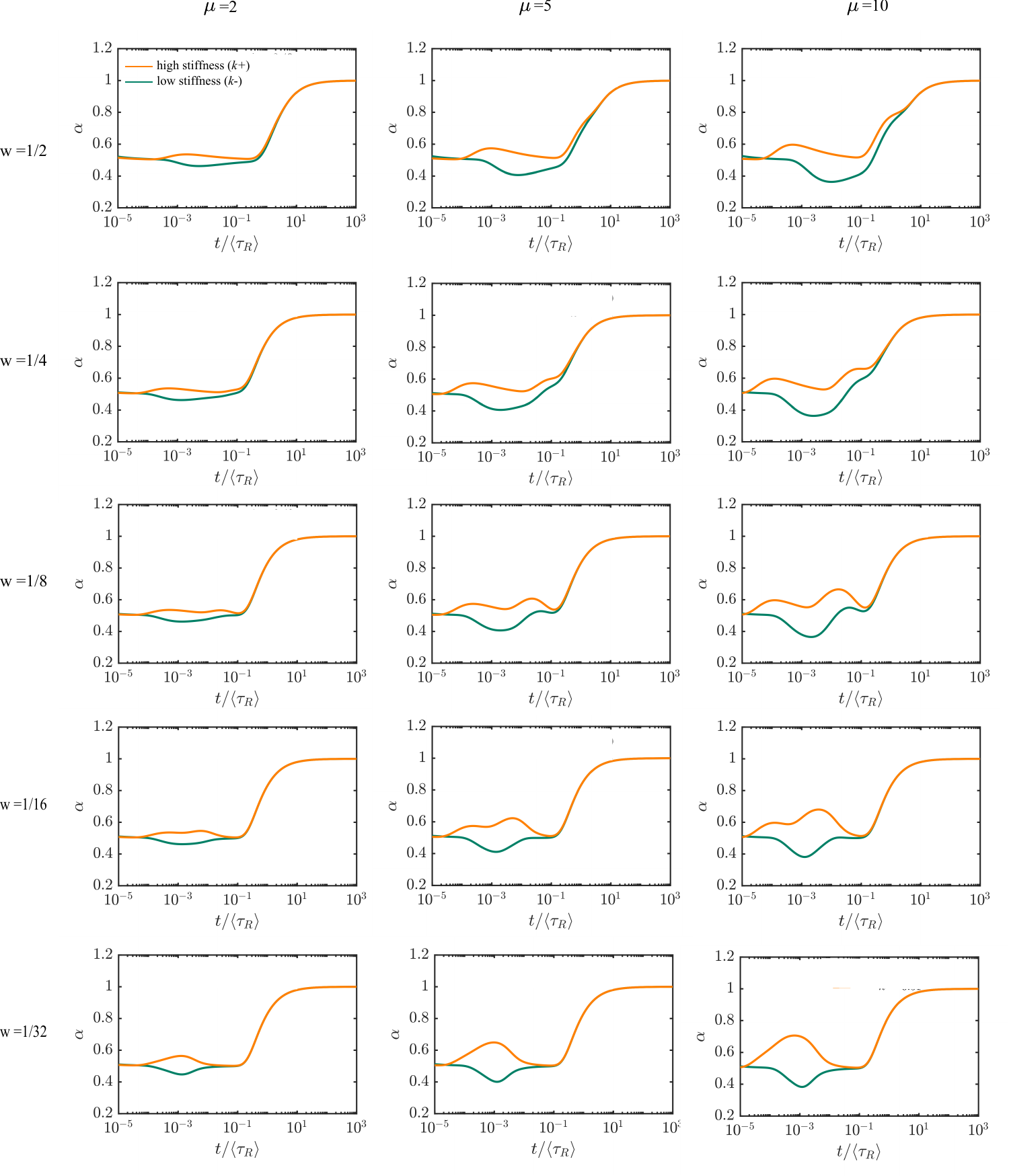}
  \caption{ Scaling exponent of MSD at different value of stiffness ratio ($\mu$) and domain width ($w$) }
  \label{fgr:example}
\end{figure*}

%\begin{figure*}[h]
%\centering
%\begin{tabular}{c}
   %\includegraphics[width=0.95\linewidth]{alpha_r_0.1304_away.png}
%   \end{tabular}
%\caption{Scaling exponent of MSD for different tagged monomers from flexible and stiff regions at \(w = 1/2\). }
%\label{fig:figure1}
%\end{figure*}

\begin{figure}[h]
\centering
\begin{tabular}{c}
   \includegraphics[]{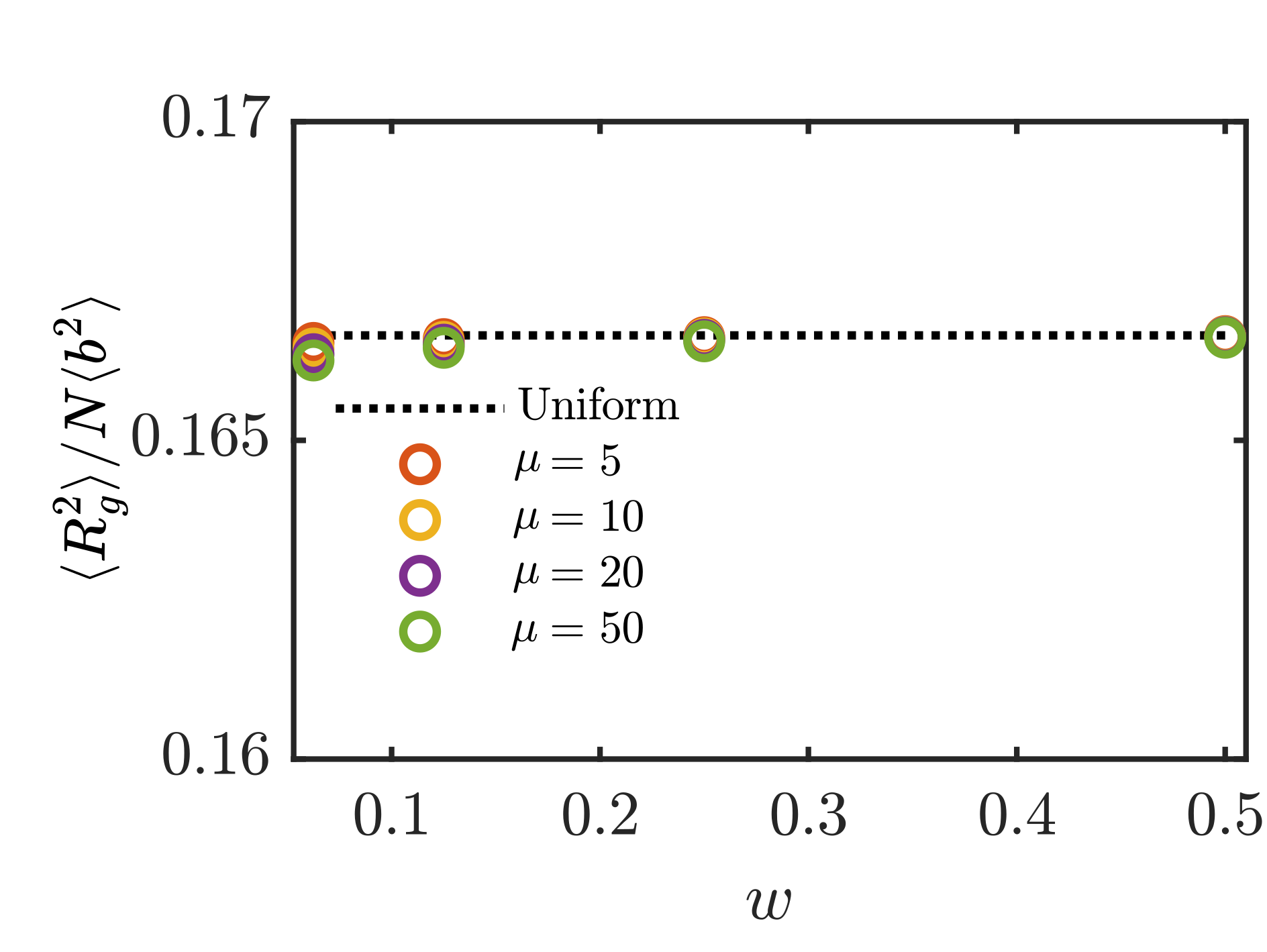}
   \end{tabular}
\caption{Mean squared radius of gyration of the polymer for different values of of stiffness ratio ($\mu$) and domain width ($w$). }
\label{fig:figure1}
\end{figure}

\begin{figure}[h]
\centering
\begin{tabular}{c}
   \includegraphics[]{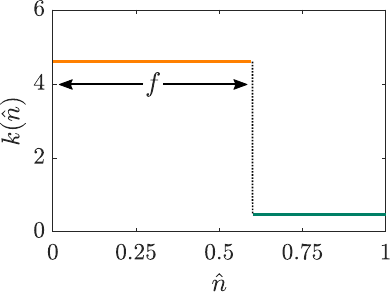}
   \end{tabular}
\caption{The stiffness of the segments as a function of segment position for polymers with unequal sized domain of high and low stiffness. $f$ represents the fraction of high stiffness domain.}
\label{fig:figure1}
\end{figure}
\end{document}